\documentclass[reprint,12pt]{elsarticle}
\usepackage{amsmath}
\usepackage[normalem]{ulem}
\usepackage{lineno,hyperref}

\newcommand{\bes}{\begin{subequations}}
\newcommand{\ees}{\end{subequations}}
\newcommand{\bea}{\begin{eqnarray}}
\newcommand{\eea}{\end{eqnarray}}
\newcommand{\ben}{\begin{equation}}
\newcommand{\een}{\end{equation}}

\journal{Physica A}

\begin{document}

\begin{frontmatter}

\title{Spin-transfer torque driven localized spin excitations in the presence of field-like torque}

\author[bdu]{M.~Lakshmanan}
\author[bdu]{R.~Arun}
\author[los]{Avadh Saxena\corref{auth1}}
\address[bdu]{Department of Nonlinear Dynamics, School of Physics, Bharathidasan University, Tiruchirappalli 620024, Tamil Nadu, India}
\address[los]{Theoretical Division and Center for Nonlinear Studies, Los Alamos National Laboratory, Los Alamos, NM 87545, USA}
\ead{avadh@lanl.gov}

\begin{abstract}
We study the existence of localized one-spin excitation in the Heisenberg one-dimensional ferromagnetic spin chain in the presence of perpendicular and parallel external magnetic fields and current with spin-transfer torque and field-like torque.  The Landau-Lifshitz-Gilbert-Slonczewski (LLGS) equation is exactly solved for the one spin excitation in the absence of onsite anisotropy for the excitations of spin with fields perpendicular and parallel to the chain.  We show the removal of damping in the spin excitations by appropriately introducing current and also the enhancement of angular frequency of the oscillations due to field-like torque in the case of both perpendicular and parallel field. The exactness of the analytical results is verified by matching with numerical counterparts.  Further, we numerically confirm the existence of in-phase and anti-phase stable synchronized oscillations for two spin-excitations in the presence of current with perpendicular field and field-like torque. We also show that the one-spin excitation is stable against thermal noise and gets only slightly modified against thermal fluctuations. 
\end{abstract}


\end{frontmatter}


\section{Introduction}
Being an important aspect in applied magnetism \cite{Hill_2002,Lak_2011}, both from theoretical and application points of view \cite{Geo_2006,Yang_2007}, the study of dynamics of classical Heisenberg ferromagnetic spin chain with anisotropic interaction is of fundamental interest in the context of spin waves in arbitrarily shaped magnetic structures \cite{Riv_2005} and spin-transfer torque in ferromagnetic layers \cite{Li_2004}.  Although there are continuum cases which are completely integrable with soliton solutions, no discrete integrable case has been studied except for the Ishimori lattice with a modified version \cite{Ishi_1984}. Two of the present authors have identified a number of special solutions in the discrete spin chain case under various situations such as with external magnetic field and onsite anisotropy \cite{Lak_2008}.  Understanding such classes of solutions in these classical spin chain systems is one of the important areas of investigations in spin dynamics.

Due to its practical relevance \cite{Sie_1988,Zolo_2003} the occurrence of localized breathers or oscillations in ferromagnetic spin chains with suitable onsite anisotropy has been studied for several years and recently two of us and Subash have studied the excitations of one, two and three spins along with their linear stability and have obtained explicit analytical solutions for the Heisenberg anisotropic spin chain in the presence of external magnetic field with onsite anisotropy \cite{Lak_2014}.  
Also, the two of us have studied the dynamics of one/many spin excitations and the impact of spin current torque in an
anisotropic Heisenberg ferromagnetic spin chain in a constant/variable external magnetic field analytically and numerically.  We have also proved analytically that the spin current can balance the damping effect and have extended the study of such model to show that in (parity-time reversal or) PT-symmetric magnetic nanostructures the gain/loss terms are canceled by the ferromagnetic coupling which leads to spin oscillations \cite{Lak_2018}.

On the other hand the spin transfer torque has been a promising candidate, starting from its discovery by Berger \cite{Berg_1996} and Slonczewski \cite{Slon_1996} theoretically as well as experimentally, for magnetization switching and magnetization oscillations with their corresponding applications in the {\it write} operation for nanomagnetic memory storage \cite{Hoso_2005} and microwave generation \cite{Kise_2003}. The dynamics of magnetization driven by spin transfer torque can be investigated numerically and analytically by solving the governing Landau-Lifshitz-Gilbert-Slonczewski (LLGS) equation \cite{Lak_2011, Li_2003}.  The role of an additional torque, known as field-like torque \cite{Zhang:prl,Zhang:prb,Galda:16}, in magnetization has been examined recently for its fruitful outcomes such as zero field oscillations \cite{Taniguchi_2014,Guo_2015} and elimination of steady state motion in coupled spin torque oscillators \cite{Arun_2019}, respectively. While the spin transfer torque transfers spin angular momentum from a pinned layer to a free layer of the spin valve system, the field-like torque arises due to the precession of spin polarized electrons, from the pinned layer, around the free layer's magnetic moment. 
The field-like torque finds technological applications in domain wall reflection \cite{Yoon_2015} as well as in magnetization uniformity in heavy metal/ferromagnetic metal/heavy metal layer structures \cite{Luo_2019}. 

The dynamics of the Heisenberg one dimensional discrete ferromagnetic spin chain for the localized excitations of the one or more spins driven by spin-transfer torque in the presence of field-like torque has not been studied yet to the best of our knowledge.  

In this paper we analytically solve the LLGS equation along with field-like torque for the components of spin for the one-spin excitation and numerically solve for the two-spin excitations in the presence of field both parallel and perpendicular to the direction of spin chain.  Also, we show the enhancement of the angular frequency of the oscillations due to the field-like torque and anisotropy.  Further, we identify the conditions among field, current and the magnitude of field-like torque to obtain the undamped oscillations for different cases. We also confirm the stable nature of the one-spin excitations against thermal fluctuations. 

The organization of the paper is as follows.  In Sec. 2 we introduce the Hamiltonian model of the spin chain system. We solve the one-spin excitation in the presence of field in Sec. 3 and current and field-like torque in Sec. 4. In Sec. 5 we solve the general case with perpendicular field, current and field-like torque. In Sec. 6 we deduce the one-spin excitation for the case of parallel field, current and field-like torque. We briefly study the two-spin excitations in the presence of perpendicular field, current and field-like torque in Sec. 7. In Sec. 8, we discuss the influence of thermal noise on one-spin excitation. Finally, in Sec. 9 we present our main conclusions.

\section{Model for spin chain}
The Hamiltonian corresponding to the evolution of $N$ number of spins of a one-dimensional anisotropic Heisenberg ferromagnetic spin chain is given by
\begin{eqnarray}
\mathcal{H} = -\sum_{\{n\}}^N (A S_n^x S_{n+1}^x+B S_n^y S_{n+1}^y + C S_n^z S_{n+1}^z) - D\sum_n (S_n^z)^2 - {\bf H}.\sum_n {\bf S}_n,\label{hamil}
\end{eqnarray}
where $S_n^x, S_n^y$ and $S_n^z$ are the spin components of the classical unit spin vector $\vec{S}_n$, satisfying the condition
\begin{eqnarray}
(S_n^x)^2+(S_n^y)^2+(S_n^z)^2=1,~~~ n=1,2,...,N. \label{norm}
\end{eqnarray}
Here $A$, $B$ and $C$ are the exchange interaction parameters, $D$ is the onsite anisotropy parameter and ${\bf H}$ is the external magnetic field.  The Landau-Lifshitz equation of motion for the $n^{th}$ spin of the chain, specified by the Hamiltonian \eqref{hamil}, is deduced by introducing appropriate spin Poisson bracket relations as\cite{Laksh_1984}
\begin{eqnarray}
\frac{d {\bf S}_n}{dt} = {\bf S}_n \times {\bf H}_{eff} + \alpha {\bf S}_n\times ({\bf S}_n\times {\bf H}_{eff})
, ~~n=1,2,...,N, \label{llg}
\end{eqnarray}
where ${\bf H}_{eff}=-\delta \mathcal{H}/\delta {\bf S_n}$ is the effective field and $\alpha$ is the Gilbert damping parameter.

\section{One-spin excitation in the presence of perpendicular field}
Consider a one-dimensional spin chain with the excitation of one spin ${\bf S}_0$ as follows:
\begin{eqnarray}
.....(1,0,0),(1,0,0),(S_0^x,S_0^y,S_0^z),(1,0,0),(1,0,0)... \,.
\end{eqnarray}
The Hamiltonian for this system, with external field ${\bf H}=(0,0,H)$ along $z$ direction, is written from Eq.\eqref{hamil} as
\begin{eqnarray}
\mathcal{H} = -[(N-3)A+2A S_0^x]-D(S_0^z)^2-H S_0^z. \label{hamil_one}
\end{eqnarray}
The effective field is derived from the Hamiltonian given in Eq.\eqref{hamil_one} as
\begin{eqnarray}
{\bf H}_{eff}=2A ~{\bf\hat i} + [2 D S_0^z+H]~{\bf\hat k},\label{heff_one}
\end{eqnarray}
where ${\bf\hat i}$ and ${\bf\hat k}$ are unit vectors along positive $x$ and $z$ directions, respectively. By substituting Eq.\eqref{heff_one} in Eq.\eqref{llg}, the equations of motion for the excited spin are obtained as
\begin{eqnarray}
\frac{dS_0^x}{dt}&=&2 D S_0^y S_0^z + H S_0^y+\alpha\left[-2A(1-(S_0^x)^2)+2 D S_0^x (S_0^z)^2+H S_0^x S_0^z\right], \label{ds0xdt}\\
\frac{dS_0^y}{dt}&=&2A S_0^z-2 D S_0^x S_0^z - H S_0^x+\alpha\left[2AS_0^x S_0^y+2 D S_0^y (S_0^z)^2+H S_0^y S_0^z\right], \label{ds0ydt}\\
\frac{dS_0^z}{dt}&=&-2AS_0^y+\alpha\left[2AS_0^x S_0^z-2D S_0^z(1-(S_0^z)^2)-H (1-(S_0^z)^2)\right]. \label{ds0zdt}
\end{eqnarray}
From Eqs.\eqref{ds0xdt},\eqref{ds0ydt} and \eqref{ds0zdt}, one can verify that 
\begin{eqnarray}
S_0^x \frac{dS_0^x}{dt}+S_0^y \frac{dS_0^y}{dt}+S_0^z \frac{dS_0^z}{dt}=0, \label{normcond}
\end{eqnarray}
to confirm that $S^2 = (S_0^x)^2+(S_0^y)^2+(S_0^z)^2= \rm{constant}=1$ is conserved. By considering the case where the onsite anisotropy is zero, we can write the dynamical equations from Eqs.\eqref{ds0xdt},\eqref{ds0ydt} and \eqref{ds0zdt}, for the one-spin excitation as
\begin{subequations}
\label{ds0dt_D0}
\begin{align}
\frac{dS_0^x}{dt}&=& H S_0^y+\alpha\left[-2A(1-(S_0^x)^2)+H S_0^x S_0^z\right], \label{ds0xdt_D0}\\
\frac{dS_0^y}{dt}&=&2A S_0^z - H S_0^x+\alpha\left[2AS_0^x S_0^y+H S_0^y S_0^z\right], \label{ds0ydt_D0}\\
\frac{dS_0^z}{dt}&=&-2AS_0^y+\alpha\left[2AS_0^x S_0^z-H (1-(S_0^z)^2)\right]. \label{ds0zdt_D0}
\end{align}
\end{subequations}

\begin{figure}[h]
	\centering\includegraphics[width=1\linewidth]{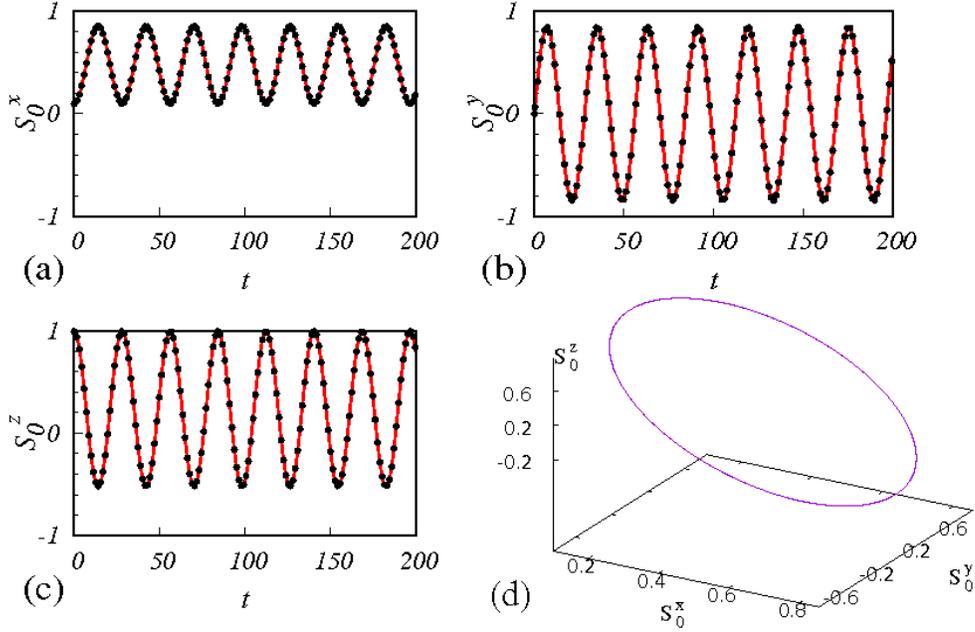}
		\caption{(Color online) Undamped oscillations of (a) $S_0^x$, (b) $S_0^y$ and (c) $S_0^z$ for $A=0.1,D=0,H=0.1$ and $\alpha=0$. Here the red lines and black dots are plotted from analytical (Eq.\eqref{S_D0}) and numerical (Eq.\eqref{ds0dt_D0}) results, respectively. The initial conditions are (0.10,0.00,0.99). (d) The three-dimensional trajectory of ${\bf S}_0$.}
		\label{fig1}
\end{figure}
\begin{figure}[h]
	\centering\includegraphics[width=1\linewidth]{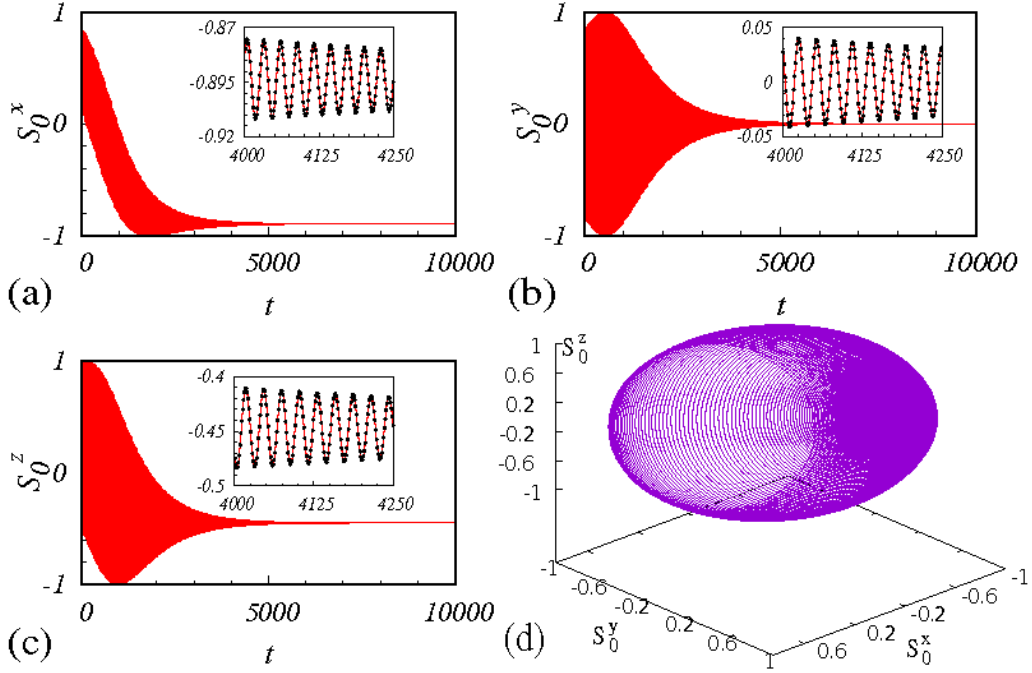}
		\caption{(Color online) Damped oscillations of (a) $S_0^x$, (b) $S_0^y$ and (c) $S_0^z$ for $A=0.1,D=0,H=0.1$ and $\alpha=0.005$ plotted from Eq.\eqref{S_D0}. In the insets for the intermediate range of time, the red lines and black dots are plotted from analytical (Eq.\eqref{S_D0}) and numerical (Eq.\eqref{ds0dt_D0}) results, respectively. The initial conditions are (0.10,0.00,0.99). (d) The three-dimensional trajectory of ${\bf S}_0$. The insets show the oscillations of the corresponding components between $t$ = 4000 and $t$ = 4250.}
	\label{fig2}
\end{figure}
Since the coupled system of Eqs.\eqref{ds0dt_D0} is difficult to solve as such, they are transformed in terms of the stereographic complex variable $\omega$ and its complex conjugate $\omega^*$ as  
\begin{eqnarray}
\omega=\frac{S_0^x+i S_0^y}{1+S_0^z},~~~~~\omega^*=\frac{S_0^x-i S_0^y}{1+S_0^z}, \label{ster1}
\end{eqnarray}
using the following transformations,
\begin{eqnarray}
S_0^x = \frac{\omega+\omega^*}{1+\omega\omega^*},~~~S_0^y = -i\frac{\omega-\omega^*}{1+\omega\omega^*},~~~S_0^z = \frac{1-\omega\omega^*}{1+\omega\omega^*}, ~~~ i=\sqrt{-1},\label{ster}
\end{eqnarray}
as follows:
\begin{eqnarray}
\frac{d\omega}{dt} = A(\alpha-i)\left[\omega^2+\frac{H}{A} \omega-1\right]=A(\alpha-i)(\omega-\omega_+)(\omega-\omega_-),\label{domegadt_Hz}
\end{eqnarray}
where $\omega_{\pm} = (1/2A)(-H\pm{\Omega})$, $\Omega=\sqrt{H^2+4A^2}$.
Eq.\eqref{domegadt_Hz} can be exactly solved as
\begin{eqnarray}
\omega(t) = \frac{\omega_+ - C ~\omega_- e^{\Omega(\alpha-i)t}}{1-C~ e^{\Omega(\alpha-i)t}},~~~\omega^*(t) = \frac{\omega_+ - C^* ~\omega_- e^{\Omega(\alpha+i)t}}{1-C^*~ e^{\Omega(\alpha+i)t}}, \label{omega_D0}
\end{eqnarray}
where $C$ and $C^*$ are arbitrary complex constants which can be obtained by using Eq.\eqref{omega_D0} as
\begin{align}
C =~ C_r+i C_i=\frac{\omega_+-\omega(0)}{\omega_--\omega(0)},~~~
C^* =~ C_r-iC_i=\frac{\omega_+-\omega^*(0)}{\omega_--\omega^*(0)}.\label{C_D0}
\end{align}
The constants $C_r$ and $C_i$ can be reexpressed using Eq.\eqref{ster1} as follows:
\begin{eqnarray}
C_{r} &=& \frac{(H^2-\Omega^2)(1+S_0^z(0))^2+4AH(1+S_0^z(0))S_0^x(0)+1-(S_0^z(0))^2}{4A^2\left(\omega_-(1+S_0^z(0)-S_0^x(0))^2+(S_0^y(0))^2\right)},\\
C_{i} &=& \frac{\Omega(1+S_0^z(0))S_0^y(0)}{A\left(\omega_-(1+S_0^z(0)-S_0^x(0))^2+(S_0^y(0))^2\right)}.
\end{eqnarray}
The components of ${\bf S}_0$ can be obtained by substituting Eqs.\eqref{omega_D0} and \eqref{C_D0} in Eqs.\eqref{ster} as
\begin{subequations}
\label{S_D0}
\begin{align}
S_0^x = \frac{2\left\{\omega_+ e^{-2\alpha\Omega t}+(H/A)e^{-\alpha\Omega t}[C_{r} \cos(\Omega t)+C_{i} \sin(\Omega t)]+\omega_-(C_{r}^2+C_{i}^2)\right\}}{(1+\omega_+^2)e^{-2\alpha\Omega t}-2(1+(H^2-\Omega^2)/4A^2)e^{-\alpha\Omega t}[C_{r}\cos(\Omega t)+C_{i}\sin(\Omega t)]+(1+\omega_-^2)(C_{r}^2+C_{i}^2)},\label{Sx_D0}\\
S_0^y = \frac{2(\Omega/A)e^{-\alpha\Omega t}[C_{r}\sin(\Omega t)-C_{i}\cos(\Omega t)]}{(1+\omega_+^2)e^{-2\alpha\Omega t}-2(1+(H^2-\Omega^2)/4A^2)e^{-\alpha\Omega t}[C_{r}\cos(\Omega t)+C_{i}\sin(\Omega t)]+(1+\omega_-^2)(C_{r}^2+C_{i}^2)},\label{Sy_D0}\\
S_0^z = \frac{(1-\omega_+^2)e^{-2\alpha\Omega t}-2(1-(H^2-\Omega^2)/4A^2)e^{-\alpha\Omega t}[C_{r}\cos(\Omega t)+C_{i}\sin(\Omega t)]+(1-\omega_-^2)(C_{r}^2+C_{i}^2)}{(1+\omega_+^2)e^{-2\alpha\Omega t}-2(1+(H^2-\Omega^2)/4A^2)e^{-\alpha\Omega t}[C_{r}\cos(\Omega t)+C_{i}\sin(\Omega t)]+(1+\omega_-^2)(C_{r}^2+C_{i}^2)}.\label{Sz_D0}
\end{align}
\end{subequations}
From Eqs.\eqref{S_D0}, we can observe that continuous oscillations are possible only in the absence of damping ($\alpha=0$).  Also, in the presence of damping, when $t\rightarrow \infty$
\begin{align}
S_0^x(\infty) = \frac{2\omega_-}{1+\omega_-^2}, ~~~
S_0^y(\infty) = 0,~~~
S_0^z(\infty) = \frac{1-\omega_-^2}{1+\omega_-^2}.\label{Sz_D0_inft}
\end{align}
Eqs.\eqref{Sz_D0_inft} clearly show that in the presence of damping the external field enables ${\bf S}_0$ to reach the steady state in the $xz$-plane. The time period of the oscillations can be determined as $T=2\pi/\Omega$.  The spin excitations of $S_0^x$, $S_0^y$ and $S_0^z$ in the absence and presence of damping are plotted in Figs.\ref{fig1} and \ref{fig2} respectively by using the expressions given in Eqs.\eqref{S_D0}.  The dots correspond to the numerical results plotted from Eqs.\eqref{ds0dt_D0}. Fig.\ref{fig1}(d) confirms the closed periodic oscillations of the components in the absence of damping. 

\section{One-spin excitation in the presence of spin-transfer torque and field-like torque}
By considering the one-dimensional spin chain in the free layer of a spin-valve (tri-layer) structure, the dynamics of the $n^{\rm th}$ spin in the presence of current is governed by the following LLGS equation\cite{Taniguchi_2014,Jerome}, 
\begin{eqnarray}
\frac{d {\bf S}_n}{dt} = {\bf S}_n \times {\bf H}_{eff} + \alpha {\bf S}_n\times ({\bf S}_n\times {\bf H}_{eff}) + j~{\bf S}_n\times({\bf S}_n\times{\bf S}_p) + j~\beta~{\bf S}_n\times{\bf S}_p, \label{llgs}
\end{eqnarray}
where $j$ is the magnitude of the { spin-transfer torque \cite{Li_2004} which can also be equivalently called the damping-like torque \cite{Manchon}} and $\beta$ is the magnitude of the field-like torque.  ${\bf S}_p=(1,0,0)$ is the polarization vector of the pinned layer.
The equations of motion of ${\bf S}_0$ in the absence of onsite anisotropy and perpendicular field are obtained by substituting Eq.\eqref{heff_one} in Eq.\eqref{llgs} as
\begin{subequations}
\label{ds0dt_H0}
\begin{align}
\frac{dS_0^x}{dt}=~& -2\alpha A_2(1-(S_0^x)^2), \label{ds0xdt_H0}\\
\frac{dS_0^y}{dt}=~& 2A_1 S_0^z +\alpha\left[2A_2S_0^x S_0^y\right], \label{ds0ydt_H0}\\
\frac{dS_0^z}{dt}=~&-2A_1S_0^y+\alpha 2A_2S_0^x S_0^z, \label{ds0zdt_H0}
\end{align}
\end{subequations}
where $A_1=A+(j\beta/2)$ and $A_2=A+(j/2\alpha)$. Eqs.\eqref{ds0dt_H0} can be transformed into stereographic form using Eqs.\eqref{ster} as follows:
\begin{eqnarray}
\frac{d\omega}{dt} &=& -i(A_1+i\alpha A_2)(\omega^2-1).\label{domegadt_H0}
\end{eqnarray}
Eq.\eqref{domegadt_H0} can be solved as
\begin{eqnarray}
\omega = \frac{1+C~e^{2(\alpha A_2-iA_1)t}}{1-C e^{2(\alpha A_2-iA_1)t}},~~~\omega^* = \frac{1+C^*~e^{2(\alpha A_2+iA_1)t}}{1-C^* e^{2(\alpha A_2+iA_1)t}},\label{omega_H0}
\end{eqnarray}
where $C=C_{r}+i C_{i}=\frac{\omega(0)-1}{\omega(0)+1}$ and $C^*=C_{r}-i C_{i}=\frac{\omega^*(0)-1}{\omega^*(0)+1}$ are arbitrary constants. $C_{r}$ and $C_{i}$ are obtained as
\begin{eqnarray}
C_{r} &=& \frac{(S_0^x(0)-S_0^z(0)-1)(S_0^x(0)+S_0^z(0)+1)+(S_0^y(0))^2}{(S_0^x(0)+S_0^z(0)+1)^2+(S_0^y(0))^2},\label{Cr_H0}\\
C_{i} &=& \frac{2S_0^y(0)(1+S_0^z(0))}{(S_0^x(0)+S_0^z(0)+1)^2+(S_0^y(0))^2}.\label{Ci_H0}
\end{eqnarray}
By substituting Eqs.\eqref{omega_H0} in Eqs.\eqref{ster} we can get the components of ${\bf S_0}$ in the presence of spin-transfer and field-like torques as
\begin{subequations}
\label{S0_H0}
\begin{align}
S_0^x &= \frac{e^{-2(2\alpha A+j)t}-(C_{r}^2+C_{i}^2)}{e^{-2(2\alpha A+j)t}+(C_{r}^2+C_{i}^2)},\label{S0xdt_H0}\\
S_0^y &= \frac{2e^{-(2\alpha A+j)t}(C_{i}\cos[(2A+j\beta)t]-C_{r}\sin[(2A+j\beta)t])}{e^{-2(2\alpha A+j)t}+(C_{r}^2+C_{i}^2)},\label{S0ydt_H0}\\
S_0^z &= \frac{-2e^{-(2\alpha A+j)t}(C_{r}\cos[(2A+j\beta)t]+C_{i}\sin[(2A+j\beta)t])}{e^{-2(2\alpha A+j)t}+(C_{r}^2+C_{i}^2)}.\label{S0zdt_H0}
\end{align}
\end{subequations}
From Eq.\eqref{S0xdt_H0} one can verify that irrespective of the field-like torque (asymptotically for large $t$) ${\bf S}_0$ approaches (-1,0,0) or (1,0,0)  when $2\alpha A+j >0$ or $2\alpha A+j<0$, respectively. Also, the current damps out the system even in the absence of damping.  Further, the periodic oscillations appear when the condition $2\alpha A+j=0$ is satisfied. From Eqs.\eqref{S0ydt_H0} and \eqref{S0zdt_H0} it can be observed that the angular frequency of the oscillations is $2A+j\beta$, which interestingly implies that the current can enhance the frequency of the oscillations only in the presence of field-like torque.

\section{Dynamics of one-spin excitation in the combined presence of perpendicular field, spin-transfer torque and field-like torque}
Equations of motion in the presence of perpendicular magnetic field ${\bf H}=(0,0,H)$, spin-transfer torque and field-like torque without onsite anisotropy can be written from Eq.\eqref{llgs} as
\begin{subequations}
\label{ds0dt_j}
\begin{align}
\frac{dS_0^x}{dt}&= H S_0^y+\alpha\left[-2
A_2(1-(S_0^x)^2)+H S_0^x S_0^z\right], \label{ds0xdt_j}\\
\frac{dS_0^y}{dt}&=2A_1 S_0^z - H S_0^x+\alpha\left[2A_2S_0^x S_0^y+H S_0^y S_0^z\right], \label{ds0ydt_j}\\
\frac{dS_0^z}{dt}&=-2A_1S_0^y+\alpha\left[2A_2S_0^x S_0^z-H (1-(S_0^z)^2)\right], \label{ds0zdt_j}
\end{align}
\end{subequations}
  \begin{figure}[h]
  	\centering
  	\includegraphics[width=1\linewidth]{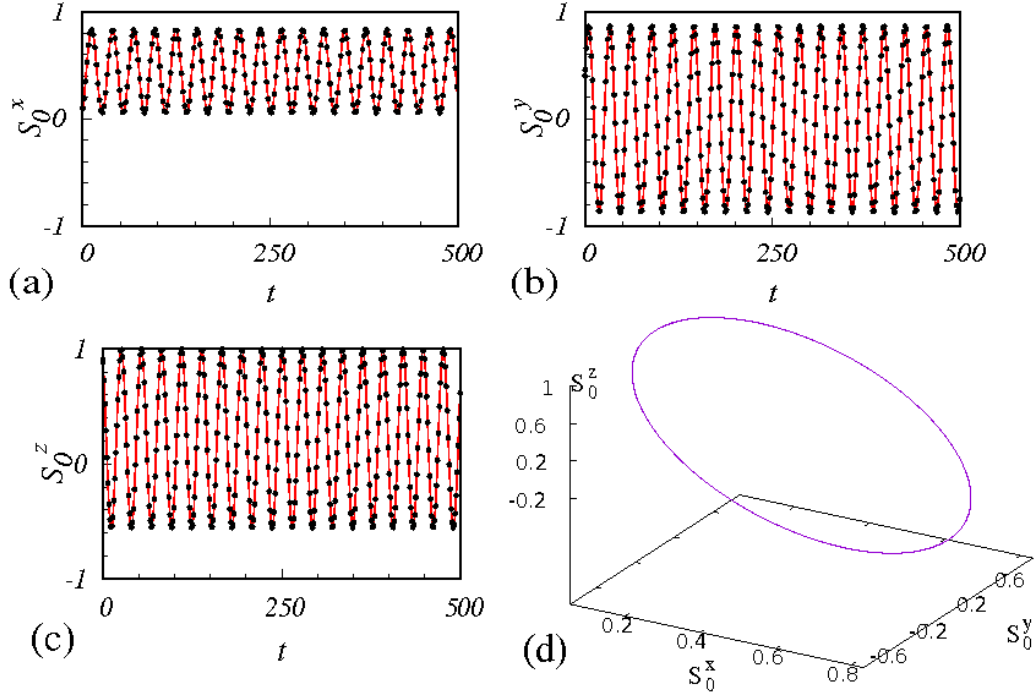}
  	\caption{(Color online) Undamped oscillations for (a) $S_0^x$, (b) $S_0^y$ and (c) $S_0^z$ when $A=0.1,D=0,~H=0.1,\alpha=0.005,~j=-0.00125016$ and $\beta=0.1$. Here the red lines and black dots are plotted from analytical (Eq.\eqref{s_j}) and numerical (Eq.\eqref{ds0dt_j}) results, respectively. The initial conditions are (0.10,0.40,0.91). (d) The three-dimensional trajectory of ${\bf S}_0$. }
  	\label{fig3}
  \end{figure}
\noindent where $A_1=A+(j\beta/2)$ and $A_2=A+(j/2\alpha)$.  Eqs.\eqref{ds0dt_j} are transformed into a stereographic equation using Eqs.\eqref{ster} as follows:
\begin{eqnarray}
\frac{d\omega}{dt} &=& (A_2 \alpha -iA_1)\left[\omega^2+ \frac{H(1+i\alpha)}{(A_1+i\alpha A_2)}\omega-1 \right]\nonumber\\
&=& (A_2\alpha-iA_1)(\omega-\omega_+)(\omega-\omega_-),\label{domegadt_j}
\end{eqnarray}
where now
\begin{align}
\omega_\pm =\left\{P_r \pm \sqrt{\frac{(Q_r^2+Q_i^2)^{1/2}}{{1+(Q_i/2Q_r)^2}}}\right\}+i\left\{ P_i\pm {(Q_i/2Q_r)} \sqrt{\frac{(Q_r^2+Q_i^2)^{1/2}}{{1+(Q_i/2Q_r)^2}}} \right\},
\end{align}
and
\begin{align}
&P_r = -\frac{H(A_2 \alpha^2+A_1)}{2(A_2^2 \alpha^2+A_1^2)},~~~P_i=-\frac{\alpha H(A_1-A_2)}{2(A_2^2\alpha^2+A_1^2)},\nonumber\\
&Q_r = \frac{H^2 (A_2\alpha^2+A_1)^2-H^2\alpha^2(A_1-A_2)^2+4(A_2^2\alpha^2+A_1^2)^2}{4(A_2^2 \alpha^2+A_1^2)},\nonumber\\
&Q_i = \frac{2\alpha H^2 (A_2\alpha^2+A_1)(A_1-A_2)}{4(A_2^2\alpha^2+A_1^2)^2}.\nonumber
\end{align}
By solving Eq.\eqref{domegadt_j}, we get
\begin{eqnarray}
\omega(t) = \frac{\omega_+-C~ \omega_- e^{(K+i\Omega)~t}}{1-C ~e^{(K+i\Omega)~t}},~~~
\omega^*(t) = \frac{\omega_+^*-C^*~ \omega_-^* e^{(K-i\Omega)~t}}{1-C^* ~e^{(K-i\Omega)~t}},\label{omega_j}
\end{eqnarray}
where $\omega_+^*$ and $\omega_-^*$ are complex conjugates of $\omega_+$ and $\omega_-$, respectively.  $C$ and $C^*$  can be derived from Eq.\eqref{omega_j} as $C =C_r+i~C_i= \frac{\omega_+-\omega(0)}{\omega_--\omega(0)}$, $C^* =C_r-i~C_i= \frac{\omega_+^*-\omega(0)^*}{\omega_-^*-\omega(0)^*}$, where $\omega(0)= \frac{S_0^x(0) + i~S_0^y}{1+S_0^z(0)}$ and $\omega(0)^*= \frac{S_0^x(0) - i~S_0^y}{1+S_0^z(0)}$. Here $K$ and $\Omega$ are given by
\begin{align}
K = 2\sqrt{\frac{(Q_r^2+Q_i^2)^{1/2}}{{1+(Q_i/2Q_r)^2}}}\left[A_2 \alpha + A_1 (Q_i/2Q_r) \right]\label{K},\\
\Omega = 2\sqrt{\frac{(Q_r^2+Q_i^2)^{1/2}}{{1+(Q_i/2Q_r)^2}}}\left[A_2 \alpha (Q_i/2Q_r) - A_1 \right].\label{Omega}
\end{align}
  \begin{figure}[h]
  	\centering
  	\includegraphics[width=0.6\linewidth]{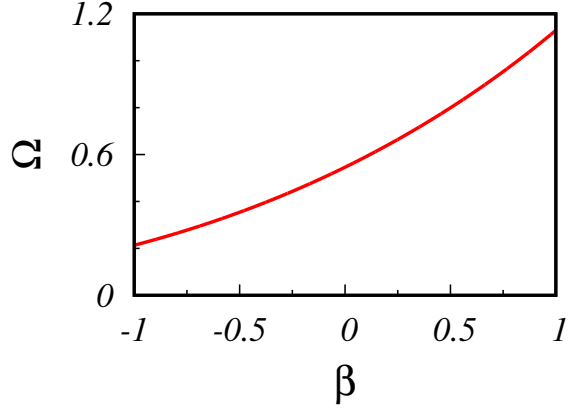}
  	\caption{(Color online) Enhancement of angular frequency of undamped oscillations by the field-like torque when $A=0.1,D=0,\alpha=0.005$ and $j=0.1$. }
  	\label{fig5}
  \end{figure}
  \begin{figure}[h]
  	\centering
  	\includegraphics[width=0.6\linewidth]{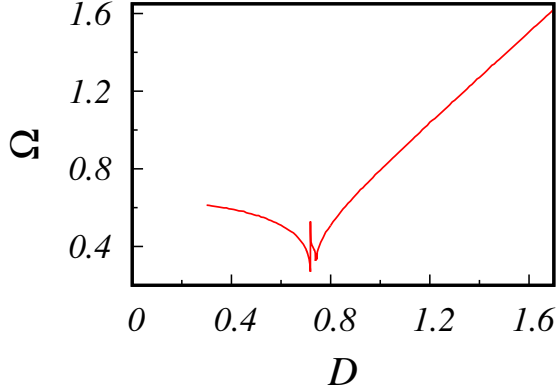}
  	\caption{(Color online) Enhancement of frequency of undamped oscillations by the anisotropy $D$ when $A=0.1,\alpha=0.005,\beta=0.1,H=0.1$ and $j=-0.01$. }
  	\label{fig5a}
  \end{figure}
From Eqs.\eqref{omega_j}, the components of ${\bf S}_0$ can be determined using Eq.\eqref{ster} as
\begin{subequations}
\label{s_j}
\begin{align}
S_0^x &= 2\left\{\frac{T_1 +T_2 e^{-2Kt} +2[T_3 \sin(\Omega t)-T_4 \cos(\Omega t)]e^{-Kt}}{T_5+T_6e^{-2Kt} +2[T_7 \sin(\Omega t)+T_8 \cos(\Omega t)]e^{-Kt}}\right\},\label{sx_j}\\
S_0^y &= 2\left\{\frac{T_9 +T_{10} e^{-2Kt} +2[T_{11} \sin(\Omega t)+T_{12} \cos(\Omega t)]e^{-Kt}}{T_5+T_6 e^{-2Kt} +2[T_7 \sin(\Omega t)+T_8 \cos(\Omega t)]e^{-Kt}}\right\},\label{sy_j}\\
S_0^z &= \left\{\frac{T_{13}+T_{14} e^{-2Kt} +2[T_{15} \sin(\Omega t)+T_{16} \cos(\Omega t)]e^{-Kt}}{T_5 +T_6e^{-2Kt} +2[T_7 \sin(\Omega t)+T_8 \cos(\Omega t)]e^{-Kt}}\right\},\label{sz_j}
\end{align}
\end{subequations}
where the explicit forms of $T_1,T_2,...,T_{16}$ are given in the Appendix.

From Eqs.\eqref{s_j} we can identify that ${\bf S}_0$ damps out and reaches steady state when $K\ne 0$.  The steady state values of $S_0^x$, $S_0^y$ and $S_0^z$ are given by 
\begin{align}
S_0^x = \frac{T_1}{T_5},~ S_0^y = \frac{T_9}{T_5}, ~S_0^z = \frac{T_{13}}{T_5},~~{\rm as}~t~{\rm tends ~to}~\infty ~{\rm when}~ K>0 ,\nonumber\\
S_0^x = \frac{T_2}{T_6}, ~S_0^y = \frac{T_{10}}{T_6}, ~S_0^z = \frac{T_{14}}{T_6},~~{\rm as}~t~{\rm  tends ~to}~\infty~{\rm when}~K<0 .\nonumber
\end{align}
When $K=0$, the undamped oscillations appear. The values of current, field and field-like torque for which undamped oscillations are possible can be obtained from Eq.\eqref{K}  as follows:
\begin{align}
&A_2 \alpha + A_1 (Q_i/2Q_r) = 0, \nonumber\\
H^2 +&\frac{4 {A_2} \left({A_1}^2+\alpha ^2 {A_2}^2\right)^2}{{A_1}^3+3 \alpha ^2 {A_1} {A_2}^2+\alpha ^4 {A_2}^3-\alpha ^2 {A_2}^3}=0.\label{cond1}
\end{align}
The angular frequency of the undamped oscillations is derived by using Eq.\eqref{cond1} in Eq.\eqref{Omega} as 
\begin{align}
\Omega = -2 \left[{\frac{({A_1}-{A_2}) \left({A_1}+\alpha ^2 {A_2}\right) \left({A_1}^2+\alpha ^2 {A_2}^2\right)\sqrt{{A_1}^2+4 \alpha ^2 {A_2}^2}}{{A_1}^3+3 \alpha ^2 {A_1} {A_2}^2+\alpha ^2 \left(\alpha ^2-1\right) {A_2}^3}}\right]^{1/2}. \label{Omega1}
\end{align}
The existence of undamped oscillations is confirmed by plotting $S_0^x$, $S_0^y$ and $S_0^z$ in Figs.\ref{fig3}(a), (b) and (c) respectively when $H=0.1,~\beta=0.1$ and $j=-0.00125$.  Also, Fig.\ref{fig3}(d) shows the three-dimensional trajectory of ${\bf S}_0$ of the undamped oscillations. The enhancement of angular frequency of undamped oscillations by field-like torque, plotted from Eq.\eqref{Omega1}, is shown in Fig.\ref{fig5} when $A=0.1,D=0,\alpha=0.005$ and $j=0.1$.  Also, the enhancement of angular frequency by the introduction of onsite anisotropy $D$ in Eqs.\eqref{ds0dt_j} is shown appropriately in Fig.\ref{fig5a} when $A=0.1,H=0.1,\alpha=0.005$, $\beta=0.1$ and $j=-0.01$, from appropriate numerical analysis.

\section{One-spin excitation in the combined presence of parallel field, spin-transfer torque and field-like torque}
When the external magnetic field is applied parallel to the chain, i.e. along x-axis, the dynamical equations can be obtained from Eq.\eqref{llgs} by considering ${\bf H}=(H,0,0)$ as
\begin{subequations}
\label{dsdt_Hx}
\begin{align}
&\frac{dS_0^x}{dt} = 2D S_0^y S_0^z -[j+\alpha(2A+H)](1-(S_0^x)^2)+2\alpha D S_0^x (S_0^z)^2,\label{dsxdt_Hx}\\
&\frac{dS_0^y}{dt} = (2A+H+j\beta)S_0^z+ [j+\alpha(2A+H)]S_0^x S_0^y+2\alpha D S_0^y(S_0^z)^2,\label{dsydt_Hx}\\
&\frac{dS_0^z}{dt} = -(2A+H+j\beta)S_0^y + [j+\alpha(2A+H)]S_0^x S_0^z - 2\alpha D S_0^z(1-(S_0^z)^2).\label{dszdt_Hx}
\end{align}
\end{subequations}
The above Eqs.\eqref{dsdt_Hx} for $D=0$ are transformed into a stereographic equation using Eq.\eqref{ster} as
\begin{align}
\frac{d\omega}{dt} = -\frac{1}{2}\left[\alpha(2A+H)+j-i(2A+H+j\beta) \right](1-\omega^2).\label{domega_Hx}
\end{align}
Eq.\eqref{domega_Hx} is solved as
\begin{align}
\omega = \frac{1+C e^{\left[j+\alpha(2A+H)-i(2A+H+j\beta) \right]t}}{1-C e^{\left[j+\alpha(2A+H)-i(2A+H+j\beta) \right]t}},~~~\omega^* = \frac{1+C^* e^{\left[j+\alpha(2A+H)+i(2A+H+j\beta) \right]t}}{1-C^* e^{\left[j+\alpha(2A+H)+i(2A+H+j\beta) \right]t}}, \label{omega_Hx}
\end{align}
where $C=C_r+iC_i=\frac{\omega(0)-1}{\omega(0)+1}$ and $C^*=C_r-iC_i=\frac{\omega^*(0)-1}{\omega^*(0)+1}$ are arbitrary constants. Here $C_{r}$ and $C_{i}$ can be obtained by using Eqs.\eqref{ster1} as in Eqs.\eqref{Cr_H0} and \eqref{Ci_H0}.
By substituting Eqs.\eqref{omega_Hx} into Eqs.\eqref{ster} we can derive 
\begin{subequations}
\label{s_Hx}
\begin{align}
&S_0^x = \frac{1-(C_{r}^2+C_{i}^2)e^{2[j+\alpha(2A+H)]t}}{1+(C_{r}^2+C_{i}^2)e^{2[j+\alpha(2A+H)]t}},\label{sx_Hx} \\
&S_0^y = 2 e^{[j+\alpha(2A+H)]t}\left\{\frac{C_{i} \cos([2A+H+j\beta]t)-C_{r} \sin([2A+H+j\beta]t)}{1+(C_{r}^2+C_{i}^2)e^{2[j+\alpha(2A+H)]t}}\right\},\label{sy_Hx}\\
&S_0^z =  -2 e^{[j+\alpha(2A+H)]t}\left\{\frac{C_{r} \cos([2A+H+j\beta]t)+C_{i} \sin([2A+H+j\beta]t)}{1+(C_{r}^2+C_{i}^2)e^{2[j+\alpha(2A+H)]t}}\right\}.\label{sz_Hx}
\end{align}
\end{subequations}
From Eqs.\eqref{s_Hx}, one can understand that the spin $S_0^x$ switches asymptotically $(t\rightarrow\infty)$ to +1 or $-1$ when $j+\alpha(2A+H)<0$ or $j+\alpha(2A+H)>0$, respectively. The angular frequency of the oscillations is given by $2A+H+j \beta$, which implies that the angular frequency is independent of the current in the absence of field-like torque. Further, it can be noticed that the undamped oscillations in the presence of parallel field are possible when the condition $j+\alpha(2A+H)=0$ is satisfied.

Manipulation  of  single  electron  spin states in solids is receiving much attention for quantum computing \cite{Kane_1998,Clark_2003}, mainly for localized electron spins in solids which show long relaxation and  coherence  times  and  their  states  can be easily manipulated via microwave or radio frequency pulses \cite{Feher_1959}.  Also, single spin dynamics in a Heisenberg XXZ spin chain has been studied for a quantum transistor \cite{Marchukov_2016} and coherent manipulation of a single spin state by microwave pulses has been investigated \cite{Jelezko_2004}.

The formation of localized spin excitations in a magnetic layer is experimentally possible. It has been proved that by means of antiferromagnetic coupling a reference layer with fixed magnetization direction can be formed from an oppositely magnetized pinned layer. Thus, it is possible to form a ferromagnetic layer with fixed direction of magnetization \cite{Kubota_2013}. It has been experimentally proved that by placing a nano-contact in this fixed layer, the localized region of magnetization beneath it can be excited by passing a current \cite{Kaka_2005,Mancoff_2005}. These works demonstrate the possibility of exciting localized spins without altering the spins outside of the localized region and reducing the number of spins by reducing the cross-sectional area of the nano-contact. Thus, the spin transfer torque cannot affect the spins other than the localized spins.

\section{Two-spin excitation in the presence of perpendicular field, spin-transfer torque and field-like torque}
The studies on one-spin excitation can be extended into multi-spin excitations in general.  In this section we numerically study the two-spin excitations in the presence of perpendicular field, current and field-like torque. The case of parallel field can also be similarly analyzed. Considering the one dimensional spin chain with the excitation of two spins ${\bf S}_0$ and ${\bf S}_1$ as follows,
\begin{eqnarray}
.....(1,0,0),(1,0,0),(S_0^x,S_0^y,S_0^z),(S_1^x,S_1^y,S_1^z),(1,0,0),(1,0,0).....,
\end{eqnarray}
the Hamiltonian for this system, with perpendicular external field ${\bf H}=(0,0,H)$ along positive $z$ direction, is written from Eq.\eqref{hamil} as
\begin{eqnarray}
\mathcal{H} &=& -[(N-4)A+A S_0^x+A S_0^x S_1^x + A S_1^x + B S_0^y S_1^y + C S_0^z S_1^z]\nonumber\\&&-D(S_0^z)^2-D(S_1^z)^2-H S_0^z -H S_1^z.\label{hamil_two}
\end{eqnarray}
The corresponding effective fields for the two spins ${\bf S}_0$ and ${\bf S}_1$ can be derived as
\begin{eqnarray}
{\bf H}_{eff,S_0} = A(1+S_1^x)~{\bf\hat i} + B S_1^y ~{\bf\hat j} + [C S_1^z+H+2D S_0^z] ~{\bf\hat k}, \label{heff_S0}\\
{\bf H}_{eff,S_1} = A(1+S_0^x)~{\bf\hat i} + B S_0^y ~{\bf\hat j} + [C S_0^z+H+2D S_1^z] ~{\bf\hat k}. \label{heff_S1}
\end{eqnarray}
The LLGS equations corresponding to the spins ${\bf S}_n, ~n=0,1,$ in the presence of field and current are given by
\begin{eqnarray}
\frac{d {\bf S}_n}{dt} = {\bf S}_n \times {\bf H}_{eff,S_n} + \alpha {\bf S}_n\times ({\bf S}_n\times {\bf H}_{eff,S_n}) + j~{\bf S}_n\times({\bf S}_n\times{\bf S}_p) + j~\beta~{\bf S}_n\times{\bf S}_p. \label{LLGS}
\end{eqnarray}
\begin{figure}[h]
	\centering
	\includegraphics[width=1\linewidth]{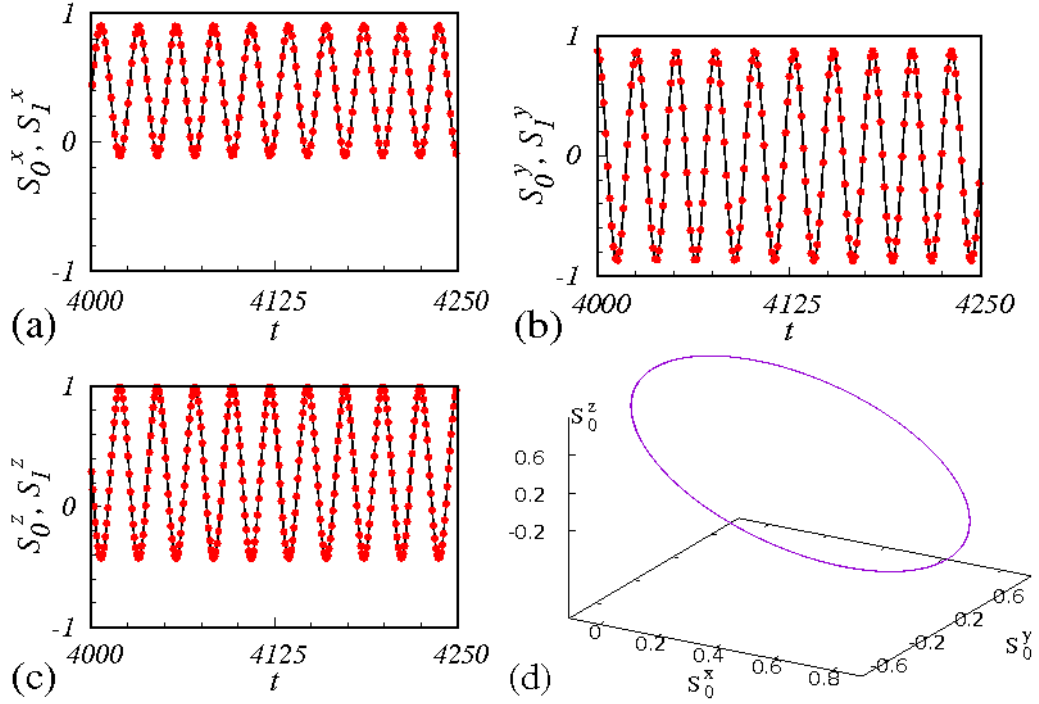}
	\caption{(Color online) Undamped oscillations of ${\bf S}_0$ (black line) and ${\bf S}_1$ (red dots). (a) $S_0^x$, $S_1^x$ (b) $S_0^y$ , $S_1^y$, (c) $S_0^z$, $S_1^z$ and (d) magnetization trajectory when $A=0.1,B=0.1,C=0.1,D=0,~H=0.1414,\alpha=0.005,~j=-0.001,~\beta=0.1$. The initial conditions are (0.6,0.8,0.0).}
	\label{fig4a}
\end{figure}
\begin{figure}[h]
  	\centering
  	\includegraphics[width=1\linewidth]{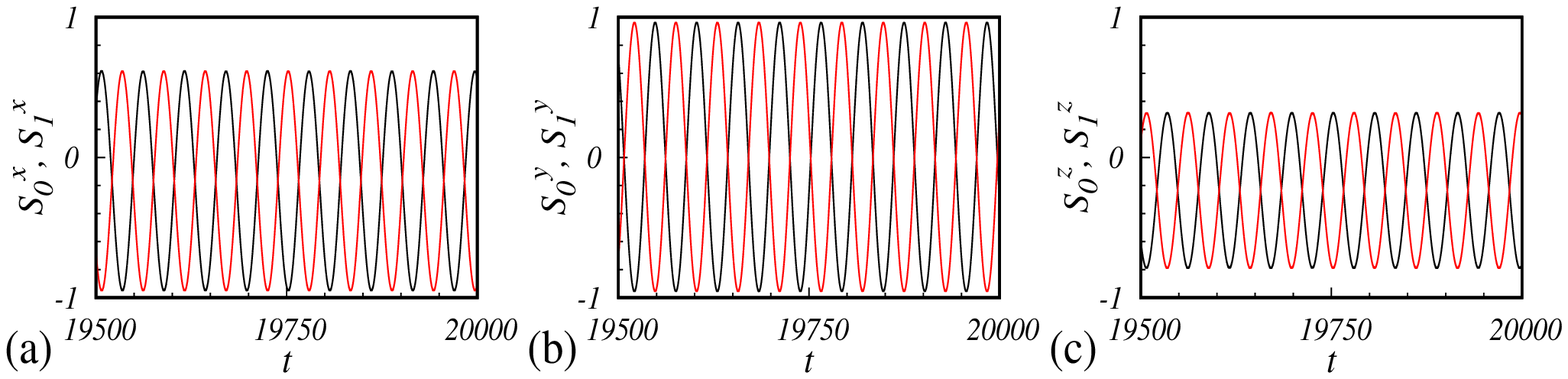}
	\caption{(Color online) Anti-phase synchronized oscillations of ${\bf S}_0$ (black line) and ${\bf S}_1$(red line). (a) $S_0^x$, $S_1^x$ (b) $S_0^y$ , $S_1^y$ and (c) $S_0^z$, $S_1^z$  when $A=0.1,B=0.1,C=0.1,D=0,~H=0.1414,\alpha=0.005,~j=-0.001,~\beta=0$. The initial conditions are (0.6,0.8,0.0) and (0.61,0.79,0.0).}
  	\label{fig4b}
\end{figure}
\begin{figure}[h]
\centering
\includegraphics[width=1\linewidth]{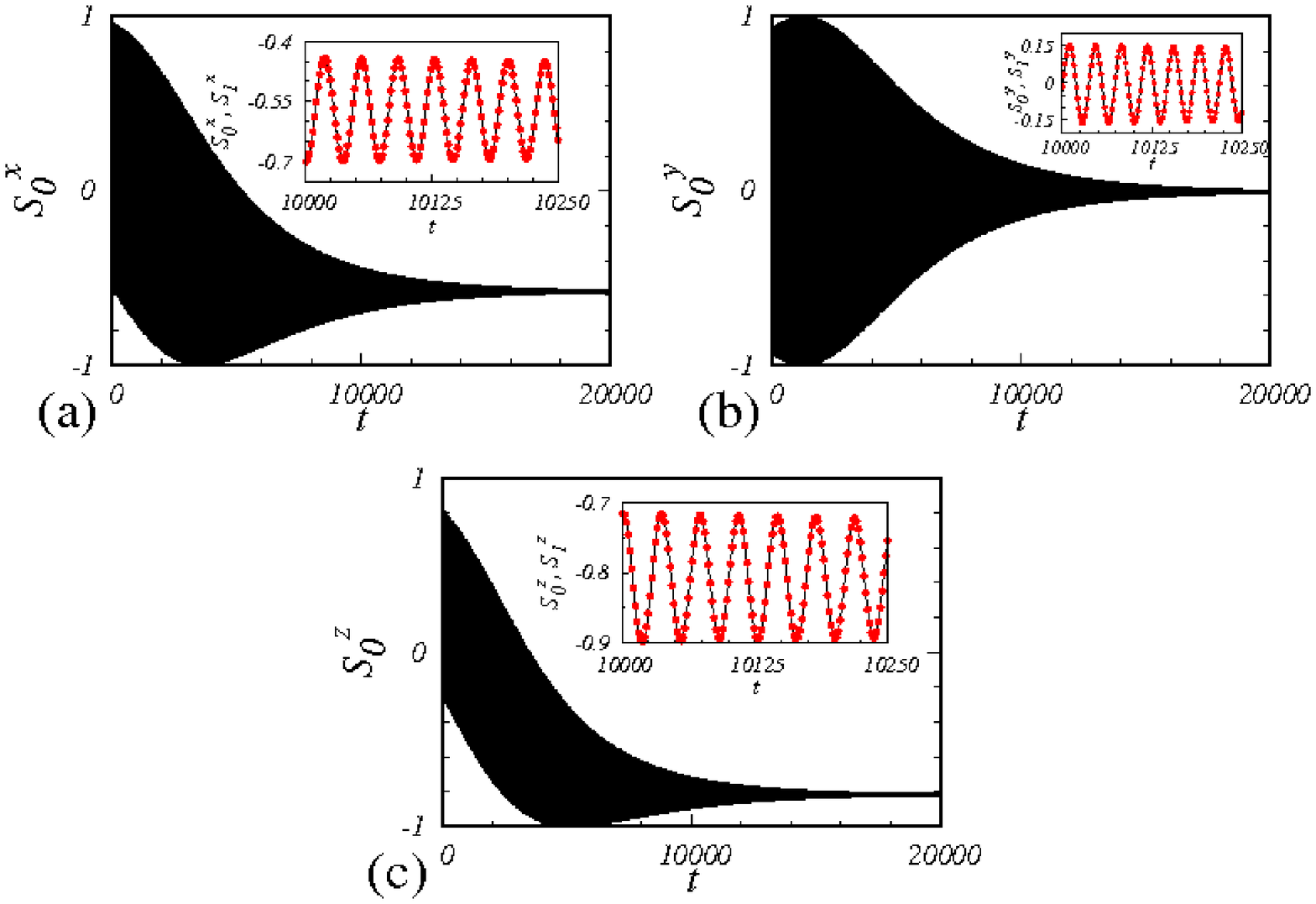}
\caption{(Color online) Damped oscillations of (a) $S_0^x$ (b) $S_0^y$ and  (c) $S_0^z$ for $A=0.1,B=0.1,C=0.1,D=0,~H=0.1414,\alpha=0.005,~j=-0.001,~\beta=0$. The initial conditions for the two spins are (0.6,0.8,0.0). The insets for the intermediate range of time show synchronization of the respective components of the two spins.}
\label{fig4c}
\end{figure}
    
The corresponding dynamical equations for the components of ${\bf S}_0$ and ${\bf S}_1$ with ${\bf S}_p=(1,0,0)$ can be derived as
\begin{subequations}
\label{2ds0dt}
\begin{align}
\frac{dS_0^x}{dt} =& C S_0^y S_1^z + 2D S_0^y S_0^z + H S_0^y - B S_0^z S_1^y \nonumber\\
&+ \alpha[-A(1-(S_0^x)^2)(1+S_1^x)+BS_0^x S_0^y S_1^y+CS_0^x S_0^z S_1^z+2D S_0^x (S_0^y)^2 +H S_0^x S_0^z]\nonumber\\&-j(1-(S_0^x)^2),\label{2ds0xdt}\\
\frac{dS_0^y}{dt} =& A S_0^z(1+S_1^x)-C S_0^x S_1^z-2D S_0^x S_0^z-H S_0^x\nonumber\\
&+\alpha[AS_0^x S_0^y(1+S_1^x)-B S_1^y (1-(S_0^y)^2)+C S_0^y S_0^z S_1^z+2D S_0^y(S_0^z)^2+HS_0^y S_0^z]\nonumber\\
&+j S_0^x S_0^y + j\beta S_0^z,\label{2ds0ydt}\\
\frac{dS_0^z}{dt} =& B S_0^x S_1^y - A S_0^y(1+S_1^x)\nonumber\\
&+ \alpha[A S_0^x S_0^z(1+S_1^x)+B S_0^y S_0^z S_1^y-(CS_1^z+2DS_0^z+H)(1-(S_0^z)^2) ]\nonumber\\
&+j S_0^x S_0^z - j\beta S_0^y,\label{2ds0zdt}
\end{align}
\end{subequations}
\begin{subequations}
\label{2ds1dt}
\begin{align}
\frac{dS_1^x}{dt} =& C S_1^y S_0^z + 2D S_1^y S_1^z + H S_1^y - B S_1^z S_0^y \nonumber\\
&+ \alpha[-A(1-(S_1^x)^2)(1+S_0^x)+BS_1^x S_1^y S_0^y+CS_1^x S_1^z S_0^z+2D S_1^x (S_1^y)^2 +H S_1^x S_1^z]\nonumber\\&-j(1-(S_1^x)^2),\label{2ds1xdt}\\
\frac{dS_1^y}{dt} =& A S_1^z(1+S_0^x)-C S_1^x S_0^z-2D S_1^x S_1^z-H S_1^x\nonumber\\
&+\alpha[AS_1^x S_1^y(1+S_0^x)-B S_0^y (1-(S_1^y)^2)+C S_1^y S_1^z S_0^z+2D S_1^y(S_1^z)^2+HS_1^y S_1^z]\nonumber\\
&+j S_1^x S_1^y + j\beta S_1^z,\label{2ds1ydt}\\
\frac{dS_1^z}{dt} =& B S_1^x S_0^y - A S_1^y(1+S_0^x)\nonumber\\
&+ \alpha[A S_1^x S_1^z(1+S_0^x)+B S_1^y S_1^z S_0^y-(CS_0^z+2DS_1^z+H)(1-(S_1^z)^2) ]\nonumber\\
&+j S_1^x S_1^z - j\beta S_1^y. \label{2ds1zdt}
\end{align}
\end{subequations}
Eqs.\eqref{2ds0dt} and \eqref{2ds1dt} for the case $D=0$ can be transformed into the stereographic form as
\begin{subequations}
\label{dometadt_2spin}
\begin{eqnarray}
\frac{d\omega_0}{dt} =&& -\frac{A}{2}(\alpha-i)(1-\omega_0^2)\left(1+\frac{\omega_1+\omega_1^*}{1+\omega_1\omega_1^*} \right)
-\frac{B}{2}(\alpha-i)(1+\omega_0^2)\left(\frac{\omega_1-\omega_1^*}{1+\omega_1\omega_1^*} \right)\nonumber\\
&&+C(\alpha-i)\omega_0\left(\frac{1-\omega_1\omega_1^*}{1+\omega_1\omega_1^*} \right)+H(\alpha-i)\omega_0, \\
\frac{d\omega_1}{dt} =&& -\frac{A}{2}(\alpha-i)(1-\omega_1^2)\left(1+\frac{\omega_0+\omega_0^*}{1+\omega_0\omega_0^*} \right)
-\frac{B}{2}(\alpha-i)(1+\omega_1^2)\left(\frac{\omega_0-\omega_0^*}{1+\omega_0\omega_0^*} \right)\nonumber\\
&&+C(\alpha-i)\omega_1\left(\frac{1-\omega_0\omega_0^*}{1+\omega_0\omega_0^*} \right)+H(\alpha-i)\omega_1 .
\end{eqnarray}
\end{subequations}

Eqs.\eqref{2ds0dt} and \eqref{2ds1dt} are numerically solved and the undamped in-phase synchronized oscillations of spins ${\bf S}_0$ and ${\bf S}_1$ are plotted in Figs.\ref{fig4a} when $A=0.1,B=0.1,C=0.1,D=0,~H=0.1414,\alpha=0.005,~j=-0.001,~\beta=0.1$ for the same initial conditions (0.6,0.8,0.0). Interestingly, the two-spin system shows  anti-phase synchronized oscillations when the initial conditions are slightly different.  Figs.\ref{fig4b} show the undamped anti-phase synchronized oscillations of spins ${\bf S}_0$ and ${\bf S}_1$  when $A=0.1,B=0.1,C=0.1,D=0,~H=0.1414,\alpha=0.005,~j=-0.001,~\beta=0.1$ for the different initial conditions (0.6,0.8,0.0) and (0.61,0.79,0.0). Damped oscillations in the absence of field-like torque are shown in Figs.\ref{fig4c}. Same results are obtained by solving the system \eqref{dometadt_2spin} as well.

\section{Effect of thermal noise on one-spin excitation in the presence of perpendicular field}
 \begin{figure}[h]
  	\centering
  	\includegraphics[width=1\linewidth]{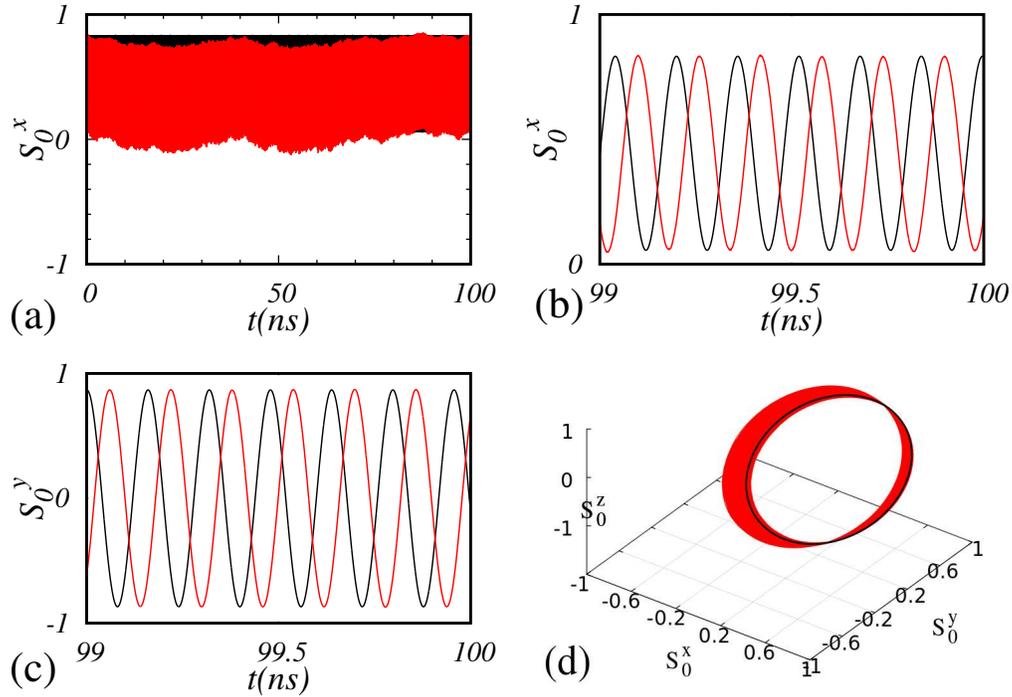}
  	\caption{ (Color online) Numerically plotted temporal evolutions of (a) \& (b) $S_0^x$ and (c) $S_0^y$ when $A=1000$ Oe, $D$=0,~$H=1000$ Oe (perpendicular field), $\alpha$=0.005,~$j$=-12.5016 (-0.1043 mA) and $\beta=0.1$. Here the red and black lines are plotted in the presence ($T$ = 300 K) and absence ($T$ = 0 K) of thermal noise, respectively. The initial conditions are (0.10,0.40,0.91). (d) The three-dimensional trajectory of ${\bf S}_0$ with (red line) and without (black line) the thermal noise. }
  	\label{fig9}
  \end{figure}
 We also investigate now the effect of thermal fluctuations on the dynamics of one-spin excitation. It is carried out by including the thermal field due to thermal noise in the effective field as follows \cite{Arun_2019}:
\begin{eqnarray}
{\bf H}_{eff}=2A ~{\bf\hat i} + [2 D S_0^z+H]~{\bf\hat k} + {\bf H}_{th},\label{heff_noise}
\end{eqnarray}
where the thermal field is given by
\begin{eqnarray}
{\bf H}_{th} = \sqrt{F} ~{\bf G},~~~F = \frac{2\alpha k_B T}{(1+\alpha^2) M_s \mu_0 V \triangle t} \,. 
\end{eqnarray}
In the above equation, ${\bf G}$ is the Gaussian random number generator vector of the oscillator with components $(G_{x} , G_{y} , G_{z} )$, which satisfies the statistical properties $\langle G_{m} (t) \rangle= 0$ and $\langle G_{m} (t) G_{n} (t') \rangle= \delta_{mn}\delta(t-t')$ for all $m, n = x, y, z$.  Here $k_B$ is the Boltzmann constant, $T$ is the temperature, $M_s$ = 1448.4 emu/cc is the saturation magnetization, $\mu_0$ is the magnetic permeability in free space, V = 2.5$\times$64$\times$64 nm$^3$~\cite{Li_2003} is a typical volume of the free layer and $\triangle t$ is the step size of the time scale used in the simulation. 

The temporal evolution of $S_0^x$ is plotted in Fig.\ref{fig9}(a) for the time range $t$ = 0 to 100 ns, where the black and red solid lines are plotted in the absence ($T$ = 0 K) and presence ($T$ = 300 K) of thermal noise, respectively, for the parameters \cite{Taniguchi_2014,Arun_2019} $A=1000$ Oe, $D$=0,~$H=1000$ Oe (perpendicular field), $\alpha$=0.005,~$j$=-12.5016 (-0.1043 mA) and $\beta=0.1$ (See Appendix B).  In Fig.\ref{fig9} we observe that there is a slight variation in the oscillation boundary without any change in the amplitude.  The smooth oscillations even in the presence of thermal noise are confirmed by plotting the temporal evolutions of $S_0^x$, $S_0^y$ and $S_0^z$ in Figs.\ref{fig9}(b), (c) and (d), respectively.  From these figures we observe that the thermal noise only very slightly affects the time evolution of the spin. 

\section{Conclusions}
By solving the LLGS equation along with field-like torque we have analytically deduced the expressions for one-spin excitation in the presence of perpendicular/parallel fields. It has been observed that the field-like torque is essential to enhance the frequency of the oscillations and it increases the frequency of oscillations for both the cases of perpendicular and parallel magnetic fields.  Relevant conditions have been obtained among the current, magnetic field and field-like torque to obtain the undamped oscillations.  The numerical study has been extended to the case of two-spin excitations and the possibility of undamped in-phase and anti-phase synchronized oscillations has been shown between the two spins. The investigations on one-spin excitation against thermal fluctuations show that the system is only slightly affected by thermal noise.  Our results are potentially important for understanding the spin dynamics in relevant magnetic materials and structures \cite{Hill_2002, Riv_2005}. 

\section*{Appendix A}
In this appendix, we provide the full expressions of various parameters $T_i, ~i$ = 1, 2, ...,16, given in Eq.\eqref{s_j}:
\begin{align}
T_1 &= (C_r^2+C_i^2) (P_r \sqrt{1+(Q_i/2Q_r)^2}-(Q_r^2+Q_i^2)^{1/4}),\nonumber\\
T_2 &= (Q_r^2+Q_i^2)^{1/4} + P_r \sqrt{1+(Q_i/2Q_r)^2},\nonumber\\
T_3 &= C_i P_r \sqrt{1+(Q_i/2Q_r)^2} - C_r (Q_r^2+Q_i^2)^{1/4} (Q_i/2Q_r),\nonumber\\
T_4 &= C_r P_r \sqrt{1+(Q_i/2Q_r)^2} + C_i (Q_r^2+Q_i^2)^{1/4} (Q_i/2Q_r),\nonumber\\
T_5 &= (C_r^2+C_i^2)\left[-2(Q_r^2+Q_i^2)^{1/4} (P_r+P_i (Q_i/2Q_r))\right.\nonumber\\&\left.+\sqrt{1+(Q_i/2Q_r)^2} (1+P_r^2+P_i^2+\sqrt{Q_r^2+Q_i^2})\right],\nonumber\\
T_6 &= 2(Q_r^2+Q_i^2)^{1/4} (P_r+P_i (Q_i/2Q_r))\nonumber\\&+\sqrt{1+(Q_i/2Q_r)^2} (1+P_r^2+P_i^2+\sqrt{Q_r^2+Q_i^2}),\nonumber\\
T_7 &= 2 C_r (Q_r^2+Q_i^2)^{1/4} (P_i-P_r(Q_i/2Q_r))\nonumber\\& + 2C_i\sqrt{1+(Q_i/2Q_r)^2} (1+P_r^2+P_i^2+\sqrt{Q_r^2+Q_i^2}),\nonumber\\
T_8 &= 2 C_i (Q_r^2+Q_i^2)^{1/4} (P_i-P_r(Q_i/2Q_r))\nonumber\\& - 2C_r\sqrt{1+(Q_i/2Q_r)^2} (1+P_r^2+P_i^2+\sqrt{Q_r^2+Q_i^2}),\nonumber\\
T_{9} &= (C_r^2+C_i^2) (P_i \sqrt{1+(Q_i/2Q_r)^2}- (Q_r^2+Q_i^2)^{1/4} (Q_i/2Q_r)),\nonumber\\
T_{10} &= (Q_r^2+Q_i^2)^{1/4} (Q_i/2Q_r)+P_i \sqrt{1+(Q_i/2Q_r)^2},\nonumber\\
T_{11} &= C_r (Q_r^2+Q_i^2)^{1/4}+C_i P_i \sqrt{1+(Q_i/2Q_r)^2},\nonumber\\
T_{12} &= C_i (Q_r^2+Q_i^2)^{1/4}-C_r P_i \sqrt{1+(Q_i/2Q_r)^2},\nonumber\\
T_{13} &= (C_r^2+C_i^2)\left[2(Q_r^2+Q_i^2)^{1/4} (P_r+P_i\sqrt{1+(Q_i/2Q_r)^2}) \right.\nonumber\\
&\left.- \sqrt{1+(Q_i/2Q_r)^2}(1-P_i^2-P_r^2-(Q_r^2+Q_i^2)^{1/4})\right],\nonumber\\
T_{14} &= -2(Q_r^2+Q_i^2)^{1/4} (P_r+P_i\sqrt{1+(Q_i/2Q_r)^2})\nonumber\\ &+ \sqrt{1+(Q_i/2Q_r)^2}(1-P_i^2-P_r^2-(Q_r^2+Q_i^2)^{1/4}),\nonumber\\
T_{15} &= 2C_r (Q_r^2+Q_i^2)^{1/4}(P_r (Q_i/2Q_r)-P_i)-C_i \sqrt{1+(Q_i/2Q_r)^2} (P_i^2+P_r^2-1-(Q_r^2+Q_i^2)),\nonumber\\
T_{16} &= 2C_r (Q_r^2+Q_i^2)^{1/4}(P_r (Q_i/2Q_r)-P_i)+C_i \sqrt{1+(Q_i/2Q_r)^2} (P_i^2+P_r^2-1-(Q_r^2+Q_i^2)).\nonumber
\end{align}

\section*{Appendix B: Comparison of numerical parameters with realistic material parameters}
Here we will briefly explain the procedure to deduce the expressions for current $j$ and the coefficient of field-like torque $\beta$ by comparing Eq. (21) with the standard form of Landau-Lifshitz-Gilbert-Slonczewski (LLGS) equation utilized for the spin torque nano oscillator (STNO) that consists of a ferromagnetic free layer and pinned layer with a nonmagnetic conducting spacer layer which separates the ferromagnetic free and pinned layers.

As discussed in Sec. 4 the LLGS equation for a spin in the presence of perpendicular field, current and field-like torque is given by
\begin{align}
\frac{d {\bf S}}{dt} = {\bf S} \times {\bf H}_{eff} + \alpha {\bf S}\times ({\bf S}\times {\bf H}_{eff}) + j~{\bf S}\times({\bf S}\times{\bf S}_p) + j~\beta~{\bf S}\times{\bf S}_p, \tag{A.1} \label{eq1}
\end{align}
where
\begin{align}
{\bf H}_{eff} = 2A ~{\bf\hat i} + H~{\bf\hat k}. \tag{A.2} \label{Heff}
\end{align}
Using the orthogonality relation ${\bf S}.\frac{d\bf S}{dt}=0$, one can deduce from Eq.\eqref{eq1} the following equation:
\begin{align}
{\bf S}\times\frac{d\bf S}{dt}~=~ {\bf S}\times({\bf S}\times{\bf H}_{eff}) - \alpha {\bf S}\times {\bf H}_{eff} - j {\bf S}\times{\bf S}_p + j\beta {\bf S}\times({\bf S}\times{\bf S}_p). \tag{A.3} \label{eq2}
\end{align}
From Eq.\eqref{eq2} we can derive,
\begin{align}
{\bf S}\times({\bf S}\times{\bf H}_{eff})~=~  {\bf S}\times\frac{d\bf S}{dt} + \alpha {\bf S}\times {\bf H}_{eff} + j {\bf S}\times{\bf S}_p - j\beta {\bf S}\times({\bf S}\times{\bf S}_p). \tag{A.4} \label{eq3}
\end{align}
By substituting Eq.\eqref{eq3} in Eq.\eqref{eq1} we obtain,
\begin{align}
\frac{d\bf S}{dt} ~=~ (1+\alpha^2) {\bf S}\times{\bf H}_{eff} + \alpha {\bf S}\times \frac{d\bf S}{dt} + j(1-\alpha\beta) {\bf S}\times({\bf S}\times{\bf S}_p) + j (\alpha+\beta) {\bf S}\times{\bf S}_p. \tag{A.5} \label{eq4}
\end{align}
With a rescaling of time $t \rightarrow -\frac{\gamma}{1+\alpha^2}t$, we get
\begin{align}
\frac{d\bf S}{dt} ~=~ -\gamma {\bf S}\times{\bf H}_{eff} + \alpha {\bf S}\times \frac{d\bf S}{dt} - \gamma j\frac{1-\alpha\beta}{1+\alpha^2} {\bf S}\times({\bf S}\times{\bf S}_p) - \gamma j \frac{\alpha+\beta}{1+\alpha^2} {\bf S}\times{\bf S}_p, \tag{A.6} \label{llgs_spin}
\end{align}
The standard form of LLGS equation used for studying the unit magnetization vector ${\bf m}$ of the free layer of the STNO is given by \cite{Taniguchi_2014,Arun_2019}
\begin{align}
\frac{d\bf m}{dt} ~=~ -\gamma {\bf m}\times{\bf H'}_{eff} + \alpha {\bf m}\times \frac{d\bf m}{dt} + \gamma H_s {\bf m}\times({\bf m}\times{\bf m}_p) - \gamma H_s \beta'~ {\bf S}\times{\bf m}_p, \tag{A.7} \label{llgs_stno}
\end{align}
where
\begin{align}
{\bf H'}_{eff} = (H_x+K_x m_x) ~{\bf\hat i} + (H_y+K_y m_y) ~{\bf\hat j} + [H_z+(K_z-N_z)m_z]~{\bf\hat k}, \tag{A.8} \label{Heff1}
\end{align}
and
\begin{align}
H_s = \frac{\hbar \eta I}{2 e M_s V}. \tag{A.9} \label{Hs}
\end{align}
Here ${\bf H'}_{eff}$ is the effective field that includes the external fields $H_x$, $H_y$ and $H_z$ along $x$, $y$ and $z$ directions, respectively, anisotropy fields $K_x$, $K_y$ and $K_z$ along $x$, $y$ and $z$ directions, respectively, and demagnetization field $N_z$ in the free layer, $\gamma$ is the gyromagnetic ratio,  $\alpha$ is the damping constant, the unit vector ${\bf m}_p$ = (1,0,0) is along the polarization of the pinned layer and $\beta'$ is field-like torque, $\hbar(=h/2\pi)$ is the reduced Planck's constant, $\eta$ is the dimensionless parameter which determines the magnitude of the spin transfer torque, $I$ is the current flowing through the free layer, $e$ is charge of the electron, $M_s$ is the saturation magnetization and $V$ is the volume of the free layer.  Here, the demagnetization field has been included only for $z$-direction since the normal of the free layer plane is along the $z$-direction.

By comparing Eqs.\eqref{llgs_spin} and \eqref{llgs_stno} we obtain the relations
\begin{align}
I ~=~ -\frac{2eM_sV j}{\hbar \eta}\left(\frac{1-\alpha\beta}{1+\alpha^2}\right),~~~~~\beta' = \frac{\alpha+\beta}{1-\alpha\beta}, \tag{A.10} \label{conversion1}
\end{align}
and similarly by comparing Eqs.\eqref{Heff} and \eqref{Heff1} we get
\begin{align}
H_x = 2A,~H_y = 0,~ H_z = H,~K_x = 0,~ K_y = 0,~K_z-N_z = 0. \tag{A.11} \label{conversion2}
\end{align}

The material parameters are adopted from Refs. \cite{Taniguchi_2014,Arun_2019}, and are given by $\alpha$ = 0.005, $|\beta'|$ $\leq$ 0.5 (which gives the condition -0.506$\le\beta\le$ 0.493), $\eta$ = 0.54, $M_s$ = 1448.3 emu/c.c., $V$ = 2.5$\times$64$\times$64~nm$^3$. To verify the impact of thermal fluctuations we have numerically plotted the temporal evolutions of ${\bf S}$ and spin trajectory using Eq.\eqref{llgs_stno} in Figs.9 with ($T$ = 300 K) and without ($T$ = 0 K) the thermal noise for the choice of the parameters \cite{Taniguchi_2014,Arun_2019} $A=1000$ Oe, $D$=0,~$H=1000$ Oe (perpendicular field), $\alpha$=0.005,~$j$=-12.5016 (I = -0.1043 mA) and $\beta=0.1$.

\section*{Acknowledgements}
The research work of ML and RA was supported by a DST-SERB Distinguished Fellowship (No.: SERB/F/6717/2017-18). ML also wishes to thank the Center for Nonlinear Studies, Los Alamos National Laboratory, USA for its warm hospitality during his visit in the summer of 2019. This work was supported in part by the U.S. Department of Energy.

\section*{References}
\bibliography{mybibfile}

\end{document}